\documentclass[aps,prc,nofootinbib,twocolumn,groupedaddress,showpacs]{revtex4}
\usepackage{float, epsfig}
\begin{document}

\title{Differential cross section for neutron-proton bremsstrahlung} 

\author{Y. Safkan}
\altaffiliation[Present Address: ]{Overteam Technologies, Istanbul, 
Turkey}
\author{T. Akdogan}
\altaffiliation[Present Address: ]{Physics Department, Bogazici 
University, Istanbul, Turkey}
\author{W. A. Franklin} 
\author{J. L. Matthews}
\author{W. M. Schmitt}
\altaffiliation[Present Address: ]{The Charles Stark Draper 
Laboratory, Inc., Cambridge, MA 02139}
\author{V. V. Zelevinsky}
\altaffiliation[Present Address: ]{19 Winchester St., 
Brookline, MA 02446} 
\affiliation{Department of Physics and Laboratory for Nuclear Science \\
Massachusetts Institute of Technology, Cambridge, Massachusetts 02139}

\author{P. A. M. Gram}
\altaffiliation[Present Address: ]{82-1021 Kaimalu Place, 
Captain Cook, HI 96704}
\author{T. N. Taddeucci}
\author{S. A. Wender}

\affiliation{Los Alamos National Laboratory, Los Alamos, New Mexico 87545}

\author{S. F. Pate}
\affiliation{Department of Physics, New Mexico State University, Las 
Cruces, New Mexico 88003}

\date{\today}

\begin{abstract}
The neutron-proton bremsstrahlung process $(np \rightarrow np\gamma)$
is known to be sensitive to meson exchange currents in
the nucleon-nucleon interaction. The triply differential cross section
for this reaction has been measured for the first time at the Los
Alamos Neutron Science Center, using an intense, pulsed beam of up to
700 MeV  neutrons to bombard a liquid hydrogen target.   Scattered
neutrons were observed at six angles between 12$^\circ$ and
32$^\circ$, and the recoil protons were observed in coincidence at
12$^\circ$, 20$^\circ$, and 28$^\circ$ on the opposite side of the
beam.  Measurement of the neutron and proton energies at known angles
allows full kinematic  reconstruction of each event.  The data are 
compared with predictions of two theoretical calculations, based on 
relativistic soft-photon and non-relativistic potential models.
\end{abstract}


\maketitle

A quantitative description of the nucleon-nucleon interaction is one
of the primary goals of nuclear physics. Elastic proton-proton and
neutron-proton scattering have been studied in detail. There is
long-standing theoretical interest in nucleon-nucleon bremsstrahlung
$(NN \rightarrow NN\gamma)$ as the simplest inelastic process in
nucleon-nucleon scattering. The final state in this reaction contains
only  one additional particle, which interacts only
electromagnetically. Bremsstrahlung probes the  
nucleon-nucleon interaction 
in a kinematic regime intermediate between elastic scattering
(angle between scattered nucleons $\theta_{NN} \simeq 90^\circ$; 
$E_\gamma$ 
= 0) and neutron-proton 
radiative capture ($\theta_{NN}$ = 0; $E_\gamma$ = maximum).  $NN$ 
bremsstrahlung necessarily involves off-shell amplitudes in the 
nucleon-nucleon 
potential, although it is now generally accepted that these are not 
directly measurable \cite{fearing}.

\begin{figure}
\epsfig{file=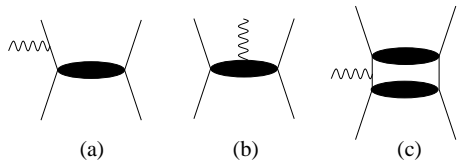,scale=0.7}
\caption{Feynman diagrams for $NN$ bremsstrahlung: (a) external
  process, (b) internal process, (c) rescattering process.}
\label{fig:feynman}
\end{figure}

The lowest-order Feynman diagrams for $NN$ bremsstrahlung are shown in
Fig. \ref{fig:feynman}: (a) the ``external'' diagram, in which the
photon is emitted by one of the nucleons\footnote{The photon 
may be emitted either before or 
after (as shown) the $NN$ interaction, and by either nucleon if both 
are charged.}, (b) the ``internal'' diagram
in which the photon is emitted during the $NN$ interaction, and (c)
the rescattering diagram. The two $NN$ bremsstrahlung reactions that
are accessible to experimental study are proton-proton bremsstrahlung
($pp\gamma$) and neutron-proton bremsstrahlung ($np\gamma$).  The
physics of $np\gamma$ differs from that of $pp\gamma$. Since the
neutron and proton can interact via the exchange of a charged meson,
the internal diagram contributes in first order.  Moreover, in
$np\gamma$ electric dipole (E1) radiation is allowed, whereas in
$pp\gamma$ the lowest allowed multipolarities are E2 and M1. As a
consequence, $pp\gamma$ cross sections are approximately an order of
magnitude smaller than $np\gamma$ cross sections.  Nonetheless, many
fewer experiments on $np\gamma$ than on $pp\gamma$ have been
attempted, owing to the difficulty of producing intense neutron beams and 
of
detecting at least one uncharged particle.

Prior to this work, no kinematically complete measurements of
$np\gamma$ cross sections had been performed. A few
doubly-differential cross section measurements, with large statistical
and systematic uncertainties \cite{Brady,Edgington}, and inclusive
photon spectrum measurements \cite{Malek,mayo} have been reported. In
the recent experiment of Volkerts {\it et al.} \cite{volkerts},
proton-neutron bremsstrahlung cross sections at incident energy 190
MeV were obtained from measurements on the quasi-free channel in
proton-deuteron bremsstrahlung ($pd \rightarrow ppn\gamma$). Events
were kinematically selected so that the second proton is likely to be
a spectator.  Although these data have excellent  statistical
precision, they cover a somewhat smaller range in outgoing nucleon
angles than in the present work and are limited to large  photon
angles.   Moreover, the  results are subject to uncertainties arising
from identifying the  quasi-free process and extracting a free $pn$
bremsstrahlung cross section from the  quasi-free cross section in a
bound system (the deuteron).

Calculations of $np\gamma$ cross sections by Brown and Franklin
\cite{vrbrown} and by Herrmann, Speth, and Nakayama \cite{herrmann}
based on non-relativistic potential models have demonstrated the
importance of the internal or meson-exchange contribution, which is seen
to increase the magnitude of the cross section by as much as a factor of
2, and to significantly alter the shape of the photon angular
distribution for relatively small nucleon emission angles.   
Similar effects are also found in the recent 
relativistic soft-photon analysis of $NN$
bremsstrahlung by Timmermans {\it et al.} \cite{timm}. The previously
measured \cite{Brady, Edgington} doubly-differential $np\gamma$ cross 
sections
$d^2\sigma /d\Omega_n d\Omega_p$ tend to exceed the calculations
\cite{vrbrown,herrmann} for larger values of $\theta_n$ and $\theta_p$.  
However, since no photon angular distributions were
extracted from these data, no stringent tests of theory could be made.

In this paper we report on the first measurement \cite{safkan} of the
triply-differential $np\gamma$ cross section  -- the photon angular 
distribution, for specific neutron and proton angles -- in the 
incident neutron energy 
region
between 175 and 275 MeV.  At the LANSCE-WNR facility at the Los 
Alamos National Laboratory, a pulsed neutron
beam  is produced by an 800-MeV proton beam incident on a 7.5-cm long
water-cooled tungsten spallation target. The proton pulse width
($<$0.5 ns) and pulse spacing (1.8 $\mu$s) allow time-of-flight
techniques to be used. The beam was defined by horizontal and vertical
copper shutters (3.8 cm x 3.8 cm opening x 40 cm long), passed through
a sweep magnet to eliminate charged particles, and then collimated by
9 m of iron to create a well-defined 2.5-cm diameter beam spot at the
$np\gamma$ target. The neutron flight path length from the spallation
source to the target is 18 m. Before reaching the target, the neutron
beam traversed a $^{238}$U fission ionization chamber \cite{wender},
in which the yield of fission fragments  was used to determine the
incident neutron flux.

The liquid hydrogen target consisted of a vertical, thin-walled (50
$\mu$m Aramica) cylindrical flask with a diameter of 7.6 cm, placed
inside a 32-cm diameter vacuum chamber. The target was operated at
a temperature of 15K and a pressure of 96.5 kPa, yielding an areal density of
0.32 atoms/barn.

Outgoing protons and neutrons were observed in two coplanar arrays of
detectors, one in the horizontal plane, and one inclined at $\sim 
25^\circ$ to 
the horizontal.  Protons were detected using pure CsI crystals,  placed 
at polar angles of $12^\circ$, $20^\circ$, and $28^\circ$ in each array. 
Each detector 
subtended a solid angle of about 8.5 msr.  Thin
plastic $\Delta E$ detectors attached to the front faces of the crystals 
provided particle identification information.  Neutron detectors,
which were liquid or plastic scintillators subtending solid 
angles of about 3 msr, were placed at polar angles of
12$^\circ$, 16$^\circ$, 20$^\circ$, 24$^\circ$, 28$^\circ$, and
32$^\circ$ in each array on the opposite side of the beam.  The front 
faces of the
neutron detectors were covered by thin plastic ``veto'' detectors, to
reject charged-particle events.  All detector gains and timing
stability were monitored continuously using light pulses from a
nitrogen laser transmitted through optical fibers to each
photomultiplier tube \cite{marshall,vzsb}.

The proton energies are determined by pulse height information from
the CsI detectors. The pulse height {\it vs.}~kinetic energy
calibration is performed using elastic $np$ scattering, which
dominates the  events recorded when the data acquisition system is
triggered by ``proton singles.'' The elastic  protons are thus easily
identified.

The arrival times of the scattered neutrons and recoil protons at
their respective detectors are measured  relative to the arrival time
of the proton pulse at the spallation source.
An independent determination of the proton
kinetic energy, based upon the pulse height measurement, makes it
possible to determine the times-of-flight
of the incident
neutrons from the spallation source to the hydrogen target and of the
scattered neutrons from the target to the detectors.

Since the maximum opening angle between the neutron and proton
detectors is 60$^\circ$, all real $np$ coincidence events must result
from inelastic scattering. Below the $\pi$-production threshold, the
only inelastic process is $np \rightarrow np\gamma$.  Candidate
bremsstrahlung events are those comprising a charged particle (a count
in both detectors of a given $\Delta E$-CsI pair) in coincidence with a
neutral particle (a count in a neutron detector but not in its
corresponding veto). The data are further reduced using proton
identification (from the $E$ {\it  vs.}~$\Delta E$ information) and
neutron identification  (from the maximum possible recoil proton pulse
height in the neutron detector for a neutron arriving with a given
time of flight)  criteria.

The experimental data must be corrected for two types of background,
that from the empty target flask and that from random
coincidences. Real $np$ coincidences can result from quasi-free
$(n,np)$ processes in the $^{12}$C and $^{16}$O nuclei in the target
flask and scattering chamber windows. Although the combined thickness
of these windows is a factor of $\sim$100 less than that of the liquid
hydrogen, the $(n,np)$ cross section is a factor of $\sim$100 larger
than that for $np\gamma$, making this background comparable to the
foreground.  Moreover, it cannot be measured simply by emptying the 
target:  the presence of the liquid hydrogen in the full target produces 
significant proton energy loss which is absent when the target is empty.  
This effect must be simulated and corrected for in the subtraction of the 
empty-target data.

Compared with the real coincidence rate from $np\gamma$, the random
coincidence rate is large for three reasons: the large $n$-$p$ elastic
scattering cross section (about three orders of magnitude greater than
that for bremsstrahlung), the large ambient neutron background, and
the (necessarily) low duty factor of the incident pulsed neutron
beam. This background was measured by combining uncorrelated
``singles'' events from the proton and neutron detectors. The result
was then normalized using the count  rate in a kinematically
impossible region and subtracted from the full-target data.

The efficiencies of the neutron detectors, which varied between
7 and 12\% over the energy range of the experiment, were measured with 
an overall uncertainty of $\sim$7\% by placing the
detectors in elastic $n$-$p$ scattering geometry.  Although the proton 
detectors are essentially 100\% efficient for charged 
particles, strong interactions in the CsI will cause some protons to be 
mis-identified.  The resulting inefficiency was determined using the 
proton singles events, which are dominated by $n$-$p$ elastic scattering.

\begin{figure} 
\epsfig{file=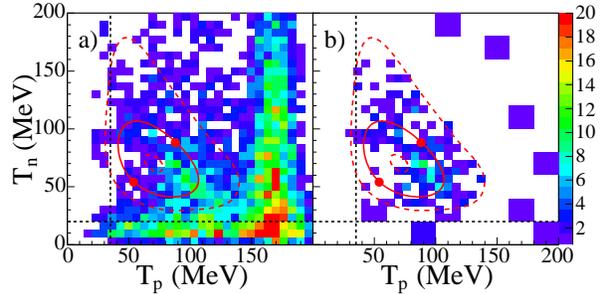,scale=0.4} 
\caption{
  (Color)
  a) Full-target data for mean incident neutron
  energy $T_{\rm inc}$ = 225 MeV and
  $\theta_n = \theta_p = 28^\circ$, plotted as a function of scattered
  neutron and proton kinetic energies. b) Data after subtraction of
  empty-target and random-coincidence background. The solid curve
  is the kinematic locus for $T_{\rm inc}$ = 225 MeV along
  which $np$ bremsstrahlung events should fall. The solid circles
  indicate  emitted photon angles of $0^\circ$ and $180^\circ$ at
  low and high nucleon energies, respectively.  The lower (upper)
  portion of the curve between $0^\circ$ and $180^\circ$ corresponds
  to photons emitted on the neutron (proton) side of the beam. The
  vertical (horizontal) short-dashed line represents the proton
  (neutron) detection threshold. The long-dashed curves illustrate a 
kinematic cut as described in the text.}
\label{fig:locus}
\end{figure}

For a given incident neutron energy, knowledge of the detector angles and
measurement of the scattered neutron and recoil proton energies uniquely
determine the energy and emission angle of the undetected photon. The
solid curve in Fig.~\ref{fig:locus}a) and b) is the kinematic locus on
which $np\gamma$ events will fall, for an incident energy of 225 MeV and
$\theta_n = \theta_p = 28^\circ$.  In Fig.~\ref{fig:locus}a) data from the
full target are plotted as a function of $T_n$ and $T_p$.  For this
presentation, data in the incident energy range 175-275 MeV are included,
with the measured neutron and proton momenta scaled to correspond to
$T_{\rm inc}$ = 225 MeV.  Fig.~\ref{fig:locus}b) shows the same data after
subtraction of the empty-target and random-coincidence backgrounds.  The
latter subtraction removes the large concentration of events seen in
Fig.~\ref{fig:locus}a) at $T_p \approx 170$ MeV; this is the recoil proton
energy in $n-p$ elastic scattering at 28$^\circ$.  To obtain the
$np\gamma$ cross section, kinematic cuts based on a Monte-Carlo analysis 
of
the finite-size effects and resolution of the experiment were applied 
to the data.  The long-dashed
curves represent the kinematic cut for $T_{\rm inc}$ = 225 MeV.
In the analysis of the data, the cuts used are a function of the 
outgoing nucleon energies and angles
{\it and} the incident beam energy.  The area enclosed by the
two long-dashed curves becomes a toroidal volume, and all the
events that lie in this volume are accepted.

In principle, the $np\gamma$ cross section could be obtained as a
function of incident neutron energy from a lower limit determined by
proton energy losses and particle detection thresholds up to the
maximum energy neutrons ($\sim$700 MeV) produced by the spallation
source. In practice, the highest incident neutron energy for which we
could measure the $np\gamma$ cross section at all angles was $\sim$300
MeV, due to the onset of the $np \rightarrow np\pi^0$ process, whose
cross section is much larger than that for $np\gamma$.  Above the
$\pi^0$-production threshold the experiment did not have adequate
resolution in missing mass to distinguish the $\gamma$ and $\pi^0$
final states.

Theoretical calculations by Brown \cite{vrbpc} and by Liou \cite{gibsonpc}
for incident neutron energies 175 and 
275 MeV showed the energy dependence of the $np\gamma$ cross section
to be much smaller than the experimental uncertainties at all 
observed nucleon 
angles.  Accordingly, in presenting the results we
have summed the cross sections obtained over this incident energy range.
For each pair of neutron and proton angles, a photon angular distribution
was constructed.  The data from each of the two coplanar arrays, after 
checking for consistency, were summed for each angle pair.

\begin{figure*}
\epsfig{file=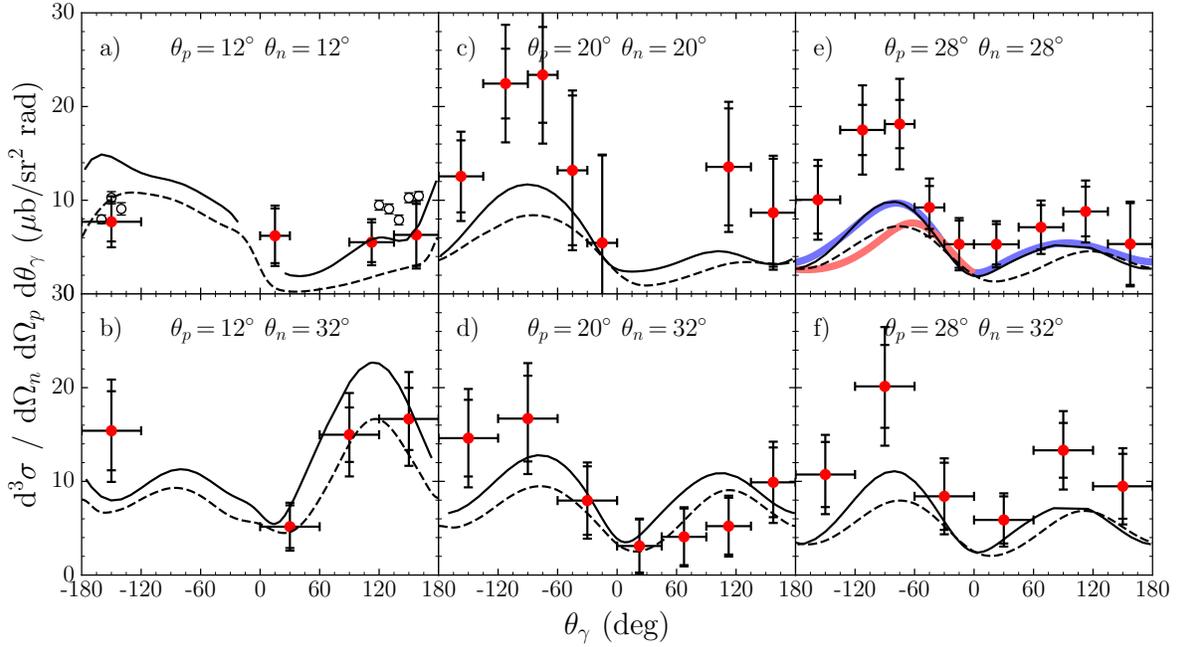,width=0.95\textwidth}
\caption{
  (Color)
  Photon angular distributions in $np$ bremsstrahlung for a) $\theta_p
  = \theta_n =  12^\circ$; b) $\theta_p = 12^\circ$, $\theta_n = 
  32^\circ$; 
  c) $\theta_p = \theta_n =  20^\circ$; d) $\theta_p = 20^\circ$,
  $\theta_n =  32^\circ$; e) $\theta_p = \theta_n = 28^\circ$; f)
  $\theta_p = 28^\circ$, $\theta_n =  32^\circ$. 
  Negative (positive) angles
  correspond to photons emitted on the neutron (proton) side of the
  beam.   The data (solid circles) have been summed over the incident
  neutron energy range 175-275 MeV; the inner error bars are the
  statistical uncertainties; the outer error bars include systematic
  uncertainties \protect\cite{syst}. The curves are the theoretical 
  predictions of Brown
  \protect\cite{vrbrown,vrbpc} (solid curves) and Timmermans  {\it et al.}
  \protect\cite{timm} (dashed curves).  The blue and red shaded bands
  represent an average over the  detector acceptances of the
  predictions of Brown  \protect\cite{vrb2006} and Timmermans  {\it et  
  al.}
  \protect\cite{tim2006}, respectively, with  out-of-plane
  contributions included.  The open circles in a) are quasi-free $pn$
  bremstrahlung cross sections at $\theta_n = 13.5^\circ$,  $\theta_p
  = 12^\circ$, and incident energy 190 MeV from the $pd$
  bremsstrahlung  experiment of Volkerts {\it et al.}
  \protect\cite{volkerts}.  The statistical uncertainties are
  comparable to the size of the symbols.
} 
\label{fig:npb}
\end{figure*}

Data were obtained during approximately six months of 
accelerator running.  The integrated beam flux in the 175-275 MeV 
energy range incident on the full and empty hydrogen target was $6 \times 
10^{12}$ and $4 \times 10^{12}$ neutrons, respectively.

As a representative sample of the data, triply differential cross sections
for six of the 18 neutron-proton angle pairs observed are shown in
Fig.~\ref{fig:npb}.  The gaps in the angular distributions for negative
(positive) angles are a consequence of the corresponding neutron (proton)
kinetic energies lying below the detection thresholds. These data and
those at the other observed angle pairs with $\theta_{n,p} \ge 20^\circ$
clearly establish an asymmetry in the photon angular distribution: the
cross section is larger for photons emitted on the neutron side than for
photons emitted on the proton side of the beam.  As is shown explicitly in
Refs. \cite{vrbrown,herrmann,timm}, such an asymmetry arises from
inclusion of meson exchange currents -- the ``internal'' diagram in Fig.
\ref{fig:feynman}(b).

The curves are results of the theoretical calculations of Brown and
Franklin \cite{vrbrown,vrbpc} and Timmermans {\it et al.} \cite{timm}.  
The agreement with the data is seen to be moderately good in most
cases;  however, there is a tendency for the theory to underpredict the
measurement.  This is particularly evident in the ``neutron-side'' region
of the angular distributions for $\theta_p = \theta_n = 20^\circ$
(Fig.~\ref{fig:npb}c)) and $\theta_p = \theta_n = 28^\circ$
(Fig.~\ref{fig:npb}e)).

The finite sizes of the proton and neutron detectors correspond to an
azimuthal (out-of-plane) angular range of about $\pm 4^\circ$, which
translates into a possible out-of-plane angle for the undetected
photon of as much as $\pm 20^\circ$.  To investigate this effect, we
have performed a Monte-Carlo average of the out-of-plane cross
sections recently calculated by Brown \cite{vrbpc,vrb2006} and by
Timmermans \cite{timm,tim2006} over the detector acceptances.  The
results are indicated by the blue and red shaded  bands in
Fig.~\ref{fig:npb}e). The average effect on the cross section of
noncoplanarity of the detected nucleons is seen to be small.

A comparison of our results with those of Volkerts {\it et al.}
\cite{volkerts}  is possible at only one set of kinematics.  In
Fig. \ref{fig:npb}a) the present  results at $\theta_n = \theta_p =
12^\circ$ are shown along with the previous results at $\theta_n =
13.5^\circ$, $\theta_p = 12^\circ$, and incident energy 190 MeV.  The
agreement between the two measurements is satisfactory within  the
experimental uncertainties, and tends to support the validity of the
renormalization factor of 0.4 which, although not understood, was
applied to the previous data \cite{volkerts}.

In summary, a measurement of the triply differential cross section for
neutron-proton bremsstrahlung has been performed over a broad kinematic
range, utilizing a pulsed neutron beam incident on a liquid hydrogen
target with coincident detection of scattered neutrons and recoil protons.
The asymmetric shapes of the measured photon angular distributions 
are consistent with those
calculated including the exchange current contribution, a 
theoretical prediction which had
not been tested experimentally.  The present results
confirm the excess of measurement over theory
suggested by previous doubly differential cross
section data, indicating that deficiencies remain in the theoretical 
models.  
The present data, albeit with limited statistical 
precision, provide 
a challenge to modern calculations of this fundamental 
process in the nucleon-nucleon interaction.

We acknowledge J. da Gra\c{c}a, M. A. Greene, and L.  W. Kwok for
their assistance in setting up the experiment and acquiring the data.
We thank V. R. Brown and B. F. Gibson for valuable guidance on 
theoretical issues and for making available their unpublished 
calculations.

This work was supported in part by funds provided by the
U.S. Department of Energy.


\begin{thebibliography}{99}
\bibitem{syst} The total systematic uncertainty is approximately 19\%, 
with the major contributions due to uncertainties in the neutron 
beam flux 
determination, the neutron detector efficiencies, and the geometrical 
acceptances of the detectors.
\end{thebibliography}
\end{document}